\documentclass{article}
\usepackage{spconf,amsmath,graphicx,bm,booktabs}

\title{Progressive multi-scale self-supervised learning for speech recognition}
\name{Genshun Wan$^{1, 2}$, Tan Liu$^2$, Hang Chen$^{1}$,Jia Pan$^2$, Cong Liu$^2$, Zhongfu Ye$^1$}
\address{$^1$University of Science and Technology of China, China \\
	$^2$iFLYTEK Research, iFLYTEK Co. Ltd., China \\
	\small{\texttt{\{gswan,ch199703\}@mail.ustc.edu.cn, \{tanliu2, jiapan,congliu2\}@iflytek.com,  yezf@ustc.edu.cn}}}
\begin{document}
%
\maketitle
\begin{abstract}
Self-supervised learning (SSL) models have  achieved considerable improvements in automatic speech recognition (ASR). In addition, ASR performance could be further improved if the model is dedicated to audio content information learning theoretically. To this end, we propose a progressive multi-scale self-supervised learning (PMS-SSL) method, which uses fine-grained target sets to compute SSL loss at top layer while uses coarse-grained target sets at intermediate layers. Furthermore, PMS-SSL introduces multi-scale structure into multi-head self-attention for better speech representation, which restricts the attention area into a large scope at higher layers while restricts the attention area into a small scope at lower layers. Experiments on Librispeech dataset indicate the effectiveness of our proposed method. Compared with HuBERT, PMS-SSL achieves 13.7\% / 12.7\% relative WER reduction on test\_other evaluation subsets respectively when fine-tuned on 10hours / 100hours subsets.

\end{abstract}
\begin{keywords}
Self-supervised learning, speech recognition, progressive, multi-scale
\end{keywords}
\section{Introduction}
\label{sec:intro}
Over the last decade, Automatic Speech Recognition (ASR) models based on Deep Neural Network (DNN) \cite{amodei2016deep,chan2021speechstew} have received increased attention and have been shown to outperform conventional models dramatically. However, the powerful modeling capability of deep structures requires a huge amount of speech-transcription pair which is expensive to collect on a large scale. Therefore, how to leverage unlabeled data to improve ASR performance is of great interest and worth exploring.Self-supervised learning (SSL) has emerged as a paradigm to learn general data representations from unlabeled examples, which has achieved impressive successes in ASR\cite{baevski2020effectiveness,baevski2019vq,baevski2020wav2vec,hsu2021hubert,liu2021tera,pascual2019learning,schneider2019wav2vec}.

For different downstream tasks, speech representation learned by SSL is expected to focus on different aspects of the spoken content, e.g., speaker identity and emotion\cite{hsu2021hubert}. Specifically, semantic content is most important for ASR.
In some successful SSL methods towards ASR, the learnt representations have high correlations with phonetic units. However, not all layers have such a high correlation. According to \cite{hsu2021hubert}, the blocks 5-12 have higher mutual information score with force-aligned phonetic transcripts than the other blocks in a 12-block HuBERT model. In order to force more layers to learn content information, \cite{wang2021self}  computes SSL loss on both intermediate layers and the top layer with the same target labels . 

We argue that hidden representations at different layers respond to the different information granularity. Specifically, the hidden representation at the higher layer may focus far more on high-level linguistic information, which is correlative with fine-grained phonetic units. The hidden representation at the lower layer is the opposite which is correlative with coarse-grained phonetic units. 

Based on the above discussion, we propose a progressive multi-scale self-supervised learning (PMS-SSL) in this paper. Specifically, we adopt the two-iteration pretraining process with masked prediction loss in HuBERT. The main contributions of this paper are listed as follows:
1)	During the second iteration, we firstly run multiple k-means clustering on the learnt representations with different cluster numbers to generate the final target sets with different granularity. Specifically, the fine-grained target sets have large cluster numbers while the coarse-grained target sets have small cluster numbers. Then we compute the SSL loss on both intermediate layers and top layer with multi tasks, where fine-grained target sets are used at top layer and coarse-grained target sets are used at intermediate layers.2)	We also find that the information granularity is closely related to the diversity and discrimination of local context and global contexts and introduce a multi-scale structure into self-attention. For the self-attention module of each layer, we restrict the attention area into different scope, which is more suited for the progressive learning strategy.

We evaluate the proposed method on Librispeech dataset. Following HuBERT, 960 hours of audio are used for pre-training and two subsets with different size (10h, 100h) are used for fine-tuning. When  progressive target sets and multi-scale self-attention are both introduced into pre-training, the model fine-tuned with the 10 hours and 100h hours subsets can achieve 13.7\% / 12.7\% word error rate (WER) reduction on test\_other subset compared with HuBERT.

\section{RELATED WORK}
\label{sec:format}

Since our proposed approach is based on HuBERT, we will give more details about HuBERT in this section.  HuBERT follows the wav2vec 2.0 architecture \cite{baevski2020wav2vec}, consisting of a convolutional waveform encoder, a BERT encoder \cite{devlin2018bert}, a projection layer and a code embedding layer. During pre-training, $p\%$of the timesteps in the output sequence of waveform encoder are randomly selected as start indices, and spans of $l$ timesteps are masked. Let $\tilde{X}=\left[\tilde{x}_1, \tilde{x}_2, \cdots \tilde{x}_T\right]$ denote the corrupted sequence of $T$ frames. The Transformer encoder consumes the corrupted sequence and outputs a hidden state sequence $O=\left[o_1, o_2, \cdots o_T\right]$, the distribution over codewords is parameterized with:
\begin{equation}
	p_f(c \mid \tilde{X}, t)=\frac{\exp \left(\operatorname{sim}\left(A o_t, e_c\right) / \tau\right)}{\sum_{c^{\prime}=1}^C \exp \left(\operatorname{sim}\left(A o_t, e_{c^{\prime}}\right) / \tau\right)}
	\label{eq1}
\end{equation}
where $A$ is the projection matrix, $e_{c}$is the embedding for codeword $c$, $sim()$ computes the cosine similarity between two vectors. A key ingredient of HuBERT is applying the prediction loss over the masked regions only, which enables the model to learn both acoustic and linguistic latent representation. We denote the cross-entropy loss computed over masked timesteps as $L_m$, which is defined as:
	\begin{equation}
		L_m=\sum_{t \in M} \log p_f(c \mid \tilde{X}, t)
		\label{eq2}
	\end{equation}
where $M$ denotes the set of timestep indices to be masked.

Following the idea that iterative refinement target labels can generate better pseudo-labels for the next iteration of training \cite{xu2020iterative,likhomanenko2020slimipl}, HuBERT conducts two iterations of pre-training. During the first iteration, HuBERT runs k-means clustering on MFCCs to generate targets for speech features. Then HuBERT re-clusters the learned latent representations and re-train the model with the newly discovered units for second iteration.
HuBERT is able to learn the speech representations which are highly correlated with phonetic units. However, not all layers have such a high correlation with phonetic units. In order to force more layers to learn more content information,\cite{wang2021self} simultaneously computes the SSL loss on both the intermediate layers and the top layer. Therefore, the total loss can be formulated as:
\begin{equation}
	L=\sum_{l \in K} L_m^l
	\label{eq3}	
\end{equation}
where $K$ denotes the set of selected layers for computing SSL loss. $L_m^l$ denotes the predictive loss at $l$-th layer, which is calculated by equation \ref{eq1} .

\section{METHOD}
\label{sec:pagestyle}
\begin{figure}
	\centering
	\includegraphics[width=1.0\linewidth]{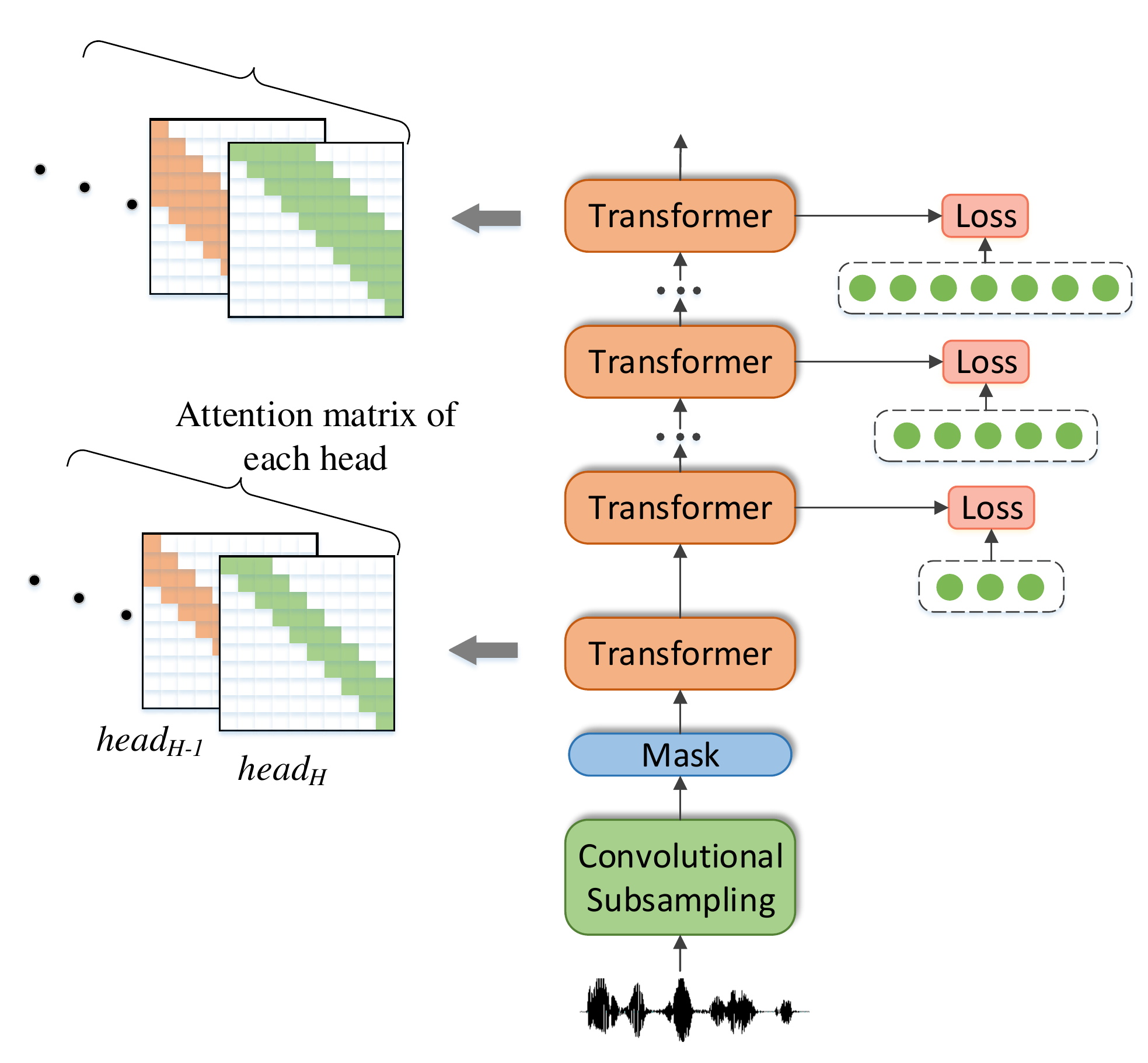}
	\caption{Model architecture}
	\label{fig:hubert}
\end{figure}

In this section, we introduce the proposed PMS-SSL method in detail. PMS-SSL assigns different-granularity target sets for different layers, and further introduces multi-scale structures into multi-head self-attention. The architecture of our model is shown in \figurename~\ref{fig:hubert}. 

\subsection{Progressive target sets}
As mentioned above, \cite{wang2021self} computes SSL loss on both intermediate layers and top layer with the same target sets to encourage more layer to learn content knowledge. However, the learnt representations at different layers  may correspond to different-granularity phonetic units. Therefore, we propose to assign different-granularity target sets for different layers. In view of the uncertainty of target sets from MFCC features, we adopt the same clustering and pre-training process as HuBERT for the first iteration. While during the second iteration, we first run multiple k-means clustering on the learnt representations with different cluster numbers to generate multiple target sets. Then we select a set of layers $K$ as the supervised layers set. Explicitly, the target sets with large size are used at top layer and the target sets with small size are used at intermediate layers. Therefore, the distribution defined in equal 1 can be refined as:
\begin{equation}
	p_f^l(c \mid \tilde{X}, t)=\frac{\exp \left(\operatorname{sim}\left(A o_t, e_c^l\right) / \tau\right)}{\sum_{c^{\prime}=1}^{c^l} \exp \left(\operatorname{sim}\left(A o_t, e_{c^{\prime}}^l\right) / \tau\right)}
\end{equation}
Where $e_c^l$ is the embedding of $c$-th label used on the $l$-th layer.

We adopt the predictive loss in HuBERT which only computes SSL loss over masked regions as the object, therefore $p_f$ in equation \ref{eq2} is replaced by $p_f^l$ and the total loss of all the supervised layers is also computed by equation \ref{eq3}.
\subsection{Multi-scale multi-head self-attention}
To obtain better speech representation suited for the progressive learning strategy, we introduce multi-scale structures into multi-head self-attention (MSA) \cite{vaswani2017attention} to emphasize the diversity of information. Multi-scale attention restricts the working area of each layer into various scopes. Intuitively, higher layers have large attention areas while lower layers have small attention areas. Compared with regular self-attention which has fixed scale of sequence length, multi-scale self-attention enable models to capture patterns at different scales and extract robust features\cite{kalchbrenner2014convolutional,chung2016hierarchical}.

Given a sequence of features $X \in R^{N \times D}$, where $N$ is the length of sequence and $D$ is the dimension of features,  regular MSA first projects $X$ into three matrices: the query $\bm{Q}$, the key $\bm{K}$ and the value $\bm{V}$, then decompose the three matrices into $H$ sub-spaces which corresponds to $H$ heads. MSA computes the scaled dot-product attention in each head and integrates the results of all heads, which can be formulated as:
\begin{equation}
	\begin{gathered}
		\operatorname{head}_i=\operatorname{softmax}\left(\frac{Q_i K_i^T}{\sqrt{d}}\right) V_i \\
		\boldsymbol{O}=\left[\text { head }_1, \text { head }_2, \cdots \text { head }_H\right] W^o
	\end{gathered}
\end{equation}
where $i$ indicates $i$-th head, $d$ indicates the dimension of each head.

As for multi-scale multi-head self-attention, the attention area of only two heads are restricted and the other heads keep the original global attention scale, which enable the model to learn both local and global context information.  A restricted attention head can be computed as:
\begin{equation}
	\operatorname{head}(X, w)_{i, j}=\operatorname{softmax}\left(\frac{Q_{i, j} C_{i, j}(K, w)^T}{\sqrt{d}} C_{i, j}(V, w)\right)
\end{equation}
where $j$ indicates $j$-th position, $C()$ is the function to restrict the attention area with a context window of length $w$ . Specially, one of the two restricted heads is used to learn history context information, thus $C()$ is defined as:
\begin{equation}
	C_{i, j}(K, w)=\left[K_{i, j-w}, \cdots, K_{i, j}\right]
\end{equation}
Another restricted heads is used to learn future context information, in which case $C()$ is formulated as:
\begin{equation}
	C_{i, j}(K, w)=\left[K_{i, j}, \cdots, K_{i, j+w}\right]
\end{equation}
The history window size and future window size at each layer are the same, while the size of window grows up progressively from bottom layer to top layer.

After pre-training, We use the connectionist temporal classification (CTC) loss \cite{graves2006connectionist} for ASR fine-tuning. Whole model weights except the convolutional audio encoder remains frozen and the projection layer is removed and replaced with a randomly initialized softmax layer. 
\section{EXPERIMENTAL DETAILS}
\label{sec:typestyle}
\subsection{Data}
We evaluated the performance of the proposed approach on
LibriSpeech task \cite{panayotov2015librispeech} . To keep the tasks comparable with the base HuBERT model, the full 960 hours of LibriSpeech audio are used for pre-training. For fine-tuning stage, we consider two different partitions: 10 hours and 100 hours. Each model is tested on the standard LibriSpeech test\_other sets and test\_clean sets respectively.

\subsection{Experiment Setup}
We adopted the pre-training process and model configure in HuBERT to train our model. To generate labels for pre-training, we ran k-means clustering with 100 clusters on 39-dimensional MFCC features over full 960 hours training audio for first iteration. As for second iteration, we first ran multiple k-means clustering with different cluster numbers $\{100, 300, 500\}$ on the latent features extracted from the 6-th transformer layer of the pre-trained model, through which we can obtain different granularity target sets. Then we computed the SSL loss on both top layer and intermediate layers with different target sets. Since the extracted feature dimension is much higher than the MFCC features, we randomly sampled 10\% of the extracted features for clustering to reduce the memory cost. For both the two iterations, the model was trained on 32 GPUs, with a batch size of at most 87.5 seconds of audio per GPU. The first iteration was trained for 250k steps, while the second iteration was trained for 400k steps.

As for the masking configurations, $p=8\%$ of the waveform encoder output frames were randomly selected as mask start and the mask span is set as $l=10$. We also used Adam optimizer with $\beta=(0.9,0.98)$ to train the model, and the learning rate ramped up linearly from 0 to the peak learning rate for the first $8\%$ of the training steps, and then decayed linearly back to zero. The peak learning rate was $5e^{-4}$. 

During the supervised fine-tuning, the model was fine-tuned on 8 GPUs with a batch size of 200 seconds of audio per GPU. In addition, the parameters of the waveform encoder were fixed, and only the new softmax matrix was trained. We used Adam optimizer with tri-stage schedule to fine-tune the model and adopted the word error rate (WER) on the dev-other subset as the criterion for model selection.

Finally, we used wav2letter++ \cite{qxjcjkgsvlrc2018wav2letter++} beam search decoder with beam size 1500 for 4-gram language model fused decoding, which optimizes:
\begin{equation}
	\log p_{C T C}(\boldsymbol{y} \mid \boldsymbol{x})+w_1 \log p_{L M}(\boldsymbol{y})+w_2|\boldsymbol{y}|
\end{equation}
where the hyperparameter settings are the same as those of HuBERT.
\subsection{Results and analysis}
Table 1 presents the performances of different pre-train models, all of which are pre-trained on 960h subset and fine-tuned on 10h and 100h subsets. We compare the proposed PMS-SSL with several competitive self-supervised approaches in the literature, including wav2vec 2.0 \cite{baevski2020wav2vec}, HuBERT\cite{ hsu2021hubert} and ILS-SSL \cite{wang2021self}. ILS-SSL adds a bucket relative position embedding to the model, which further improves the performance. For comparison with ILS-SSL, we select same supervised layer set $\{4,12\}$ and target set $\{500,500\}$ to pre-trained models without bucket relative position embedding. 

As for PMS-SSL, we set the supervised layer set K as $\{6, 12\}$, and the target sets with size $\{300, 500\}$ are used for computing SSL loss. What's more, we set $w$ (attention context window size) at layers 1-6 as 80 and set $w$ at layers 7-12 as 160. In the ultra-low resource setup with 10 hours of labeled data, PMS-SSL obtain the WERs of 4.11\% / 8.27\% on test\_clean / test\_other set, which achieves a relative word error rate reduction of  4.6\% / 13.7\% compared with HuBERT, and 3.9\% / 5.8\% compared with ISL-SSL. By further increasing the amount of labeled data to 100 hours, PMS-SSL can also get stable improvement, with a relative word error rate reduction of 5.9\% / 12.7\% compared with HuBERT, and 3.3\% / 5.3\% compared with ISL-SSL. 
\begin{table}
	\centering
	\begin{tabular}{lccc}
		\hline Model & LM & Test\_clean & Test\_other \\
		\hline \hline \textbf{\textit{10-hour labeled}} & & &  \\
		\hline wav2vec 2.0 Base & None & $11.1$ & $17.6$ \\
		HuBERT Base & None & $10.1$ & $16.8$ \\
		ILS-SSL Base & None & $10.79$ & $16.53$ \\
		wav2vec 2.0 Base & 4-gram & $4.3$ & $9.5$ \\
		HuBERT Base & 4-gram & $4.3$ & $9.4$ \\
		ILS-SSL Base & 4-gram & $4.27$ & $8.75$ \\
		\hline PMS-SSL & None & $9.43$ &$15.02$ \\
		PMS-SSL & 4-gram &$4.11$ & $8.27$ \\
		\hline \hline \textbf{\textit{100-hour labeled}} & & & \\
		\hline wav2vec 2.0 Base & None & $6.1$ & $13.3$ \\
		HuBERT Base & None & $6.3$ & $13.2$ \\
		ILS-SSL Base & None & $6.06$ & $11.89$ \\
		wav2vec 2.0 Base & 4-gram & $3.4$ & $8.0$ \\
		HuBERT Base & 4-gram & $3.4$ & $8.1$ \\
		ILS-SSL Base & 4-gram & $3.32$ & $7.59$ \\
		\hline PMS-SSL & None & $5.52$ & $11.21$\\
		PMS-SSL & 4-gram & $3.21$ & $7.19$ \\
		\hline \\

	\end{tabular}
	\caption{Results and comparison with the literature on low resource setups (10-hour, and 100-hour of labeled data).} 
\end{table}
\begin{table}
	\centering
	\begin{tabular}{ccc}
		\hline Target\_layer & Target\_label & Test\_other \\
		\hline 4,12 & 300,500 & $8.76$ \\
		6,12 & 300,500 & $\textbf{8.55}$ \\
		8,12 & 300,500 & $9.04$ \\
		$3,6,12$ & $100,300,500$ & $8.71$\\
		$4,8,12$ & $100,300,500$ & $8.83$ \\
		\hline \\
	\end{tabular}
	\caption{Results of the models with different supervised layer.} 
\end{table}

To explore how the selected supervised layer set K and the target sets influence the performance of pre-trained models, we perform some ablation experiments without multi-scale self-attention. We select different sizes of target sets for the selected supervised layer. All ablation models are finetuned with 10h subset and the results evaluated on test\_other set are listed in Table 2, where target\_layer denotes K and target\_label denotes the size of selected target sets. As shown in Table 2, supervised layer set $\{6,12\}$ paired with target set $\{300,500\}$ achieves the best performance. What’s more, both the two sets: $\{4,12\}$,$\{6,12\}$ outperform the set $\{8,12\}$, which indicates that small-size target sets are more suitable for lower layers. In addition, more supervised layers such as $\{4,8,12\}$ or $\{3,6,12\}$ cannot guarantee better performance.

Furthermore, we build some ablation models with different attention window size $w$ to explore the effect of attention scales. As mentioned in section 3.2, the attention areas of only two heads are restricted. Based on the model with supervised layer set $\{6,12\}$, we restrict the attention area of each layer into variable scopes. The results of different attention scale combinations are listed in Table 3.  The setting at first line restricts the attention area of each layer into an 80-timestep context window, which achieves 2.0\% WER reduction on test\_other. The setting at second line enables the attentions of layer 1-6 work on an 80-timestep scope and enables the attentions of layers 7-12 work on a 160-timestep scope, which obtains 3.4\% WER reduction.
\begin{table}
	\centering
	\begin{tabular}{ccc}
		\hline layer & Window\_size & Test\_other \\
    	\hline base & \  &  8.55 \\
    	1-12 & 80 &  8.38 \\
		1-6,7-12 & 80,160 & 8.27 \\
		\hline \\
	\end{tabular}
	\caption{Results of the models with different attention scale.} 
\end{table}

\section{Conclusion}
In this paper, we propose a progressive multi-scale self-supervised learning method for speech recognition. PMS-SSL assigns different granularity target sets for different layer, the target set size increase progressively from the bottom supervised layer to the top supervised layer. Furthermore, PMS-SSL introduce multi-scale structure into multi-head self-attention, which restricts the attention area into a small scope at lower layers while restricts the attention area into a large scope at higher layers. Experiments indicate that PMS-SSL outperforms the base HuBERT model.

\newpage
\bibliographystyle{IEEEbib}
\bibliography{strings,refs}

\end{document}